\documentclass[sigconf,screen]{acmart}

\AtBeginDocument{%
  }

\setcopyright{none}

\acmConference['IFL 2023]{The 35th Symposium on Implementation and Application of Functional Languages}{August 29--31,
  2023}{Braga, Portugal}

\usepackage{cleveref}
\usepackage{subcaption}
\usepackage{listings}

\usepackage{ebproof}

\makeatletter
\newcommand{\leqnomode}{\tagsleft@true\let\veqno\@@leqno}
\newcommand{\reqnomode}{\tagsleft@false\let\veqno\@@eqno}
\makeatother

\begin{document}

\title{A Frame Stack Semantics for Sequential Core Erlang}
\subtitle{Draft paper}

\author{Péter Bereczky}
\email{berpeti@inf.elte.hu}
\orcid{0000-0003-3183-0712}
\author{Dániel Horpácsi}
\orcid{0000-0003-0261-0091}
\email{daniel-h@elte.hu}
\affiliation{%
  \institution{ELTE Eötvös Loránd University}
  \city{Budapest}
  \country{Hungary}
}

\author{Simon Thompson}
\orcid{0000-0002-2350-301X}
\email{S.J.Thompson@kent.ac.uk}
\affiliation{%
  \institution{ELTE Eötvös Loránd University}
  \city{Budapest}
  \country{Hungary}
}
\affiliation{
  \institution{University of Kent}
  \city{Canterbury}
  \country{United Kingdom}
}

\renewcommand{\shortauthors}{Bereczky, Horpácsi and Thompson}

\begin{abstract}

We present a small-step, frame stack style, semantics for sequential Core Erlang, a dynamically typed, impure functional programming 
language. The semantics and the properties that we prove are machine-checked with the Coq proof assistant. We improve on previous 
work by including exceptions and exception handling, as well as built-in data types and functions. Based on the semantics, we define 
multiple concepts of program equivalence (contextual, CIU equivalence, and equivalence based on logical relations) and prove that the 
definitions are all equivalent. Using this we are able to give a correctness criterion for refactorings by means of contextually equivalent symbolic expression pairs, which is one of the main motivations of this work. 
\end{abstract}




\newcommand{\ose}{\Downarrow}
\newcommand{\hda}{\mid}
\newcommand{\bosse}[5]{\langle #1, #2, #3 \rangle \ose \{ #4, #5 \}}
\newcommand{\inlcv}[1]{\textit{inl}\ \textit{#1}}
\newcommand{\inrcv}[1]{\textit{inr}\ \textit{#1}}
\newcommand{\inl}[1]{\textit{inl}\ #1}
\newcommand{\inr}[1]{\textit{inr}\ #1}
\newcommand{\doubleplus}{\mathbin{{+}\mspace{-5mu}{+}}}
\newcommand{\concat}{\doubleplus}
\newcommand{\sub}{\sigma}
\renewcommand{\dots}{\makebox[1em][c]{.\hfil.\hfil.}}

\newcommand{\fun}[2]{\mathtt{fun}(#1) \rightarrow #2}
\newcommand{\clause}[3]{#1 \mathtt{\ when\ } #2 \rightarrow #3}
\newcommand{\case}[2]{\mathtt{case\ } #1 \mathtt{\ of\ } #2 \mathtt{\ end}}
\newcommand{\letrec}[2]{\mathtt{letrec\ } #1 \mathtt{\ in\ } #2}
\newcommand{\elet}[3]{\mathtt{let\ } #1 = #2 \mathtt{\ in\ } #3}
\newcommand{\seq}[2]{\mathtt{do\ } #1\ #2}
\newcommand{\apply}[2]{\mathtt{apply\ } #1(#2)}
\newcommand{\econs}[2]{{\normalfont \mathtt{[} #1 \mathtt{|} #2 \mathtt{]}}}
\newcommand{\vcons}[2]{{\normalfont \mathtt{[} #1 \mathtt{|} #2 \mathtt{]}}}
\newcommand{\etuple}[1]{{\normalfont \mathtt{\{} #1 \mathtt{\}}}}
\newcommand{\vtuple}[1]{{\normalfont \mathtt{\{} #1 \mathtt{\}}}}
\newcommand{\mapitem}[2]{#1 \Rightarrow #2}
\newcommand{\emap}[1]{{\normalfont \mathtt{\sim\kern-3pt\{} #1 \mathtt{\}\kern-3pt\sim}}}
\newcommand{\vmap}[1]{{\normalfont \mathtt{\sim\kern-3pt\{} #1 \mathtt{\}\kern-3pt\sim}}}
\newcommand{\clos}[3]{\textit{clos}(#1, [#2], #3)}
\newcommand{\nil}[0]{{\normalfont \mathtt{[]}}}
\newcommand{\call}[3]{\mathtt{call\ } #1\mathtt{:}#2(#3)}
\newcommand{\primop}[2]{\mathtt{primop\ } #1(#2)}
\newcommand{\try}[5]{\mathtt{try\ } #1 \mathtt{\ of\ } #2 \rightarrow #3 \mathtt{\ catch\ } #4 \rightarrow #5}
\newcommand{\evalues}[1]{\mathtt{<}#1\mathtt{>}}
\newcommand{\vvalues}[1]{\mathtt{<}#1\mathtt{>}}
\newcommand{\varlist}[1]{\mathtt{<}#1\mathtt{>}}
\newcommand{\exc}[3]{\mathtt{\{}#1, #2, #3 \mathtt{\}}^X}

\newcommand{\idfs}{\varepsilon}
\newcommand{\consfs}[2]{#1 :: #2}
\newcommand{\fparams}[2]{#1(#2)}

\newcommand{\ctxpre}{\leq_{\textit{ctx}}}
\newcommand{\ctxequiv}{\equiv_{\textit{ctx}}}
\newcommand{\ciupre}{\leq_{\textit{ciu}}}
\newcommand{\ciuequiv}{\equiv_{\textit{ciu}}}
\newcommand{\behpre}{\leq_{b}}
\newcommand{\behequiv}{\equiv_{b}}

\newcommand{\expscoped}[2]{#1 \vdash #2}
\newcommand{\redscoped}[2]{#1 \vdash #2}
\newcommand{\excscoped}[2]{#1 \vdash #2}
\newcommand{\valscoped}[2]{#1 \vdash #2}
\newcommand{\subscoped}[3]{#1 \vdash #2 \multimap #3}
\newcommand{\framesclosed}[1]{\mathtt{FSC\ } #1}
\newcommand{\frameclosed}[1]{#1 \textit{ is closed }}

\newcommand{\subst}[2]{#1[#2]}
\newcommand{\fsubst}[2]{#1[#2]}
\newcommand{\csubst}[2]{#1[#2]}
\newcommand{\update}[2]{#1\{#2\}}
\newcommand{\idsubst}{\textit{id}}
\newcommand{\restrict}{\setminus}
\newcommand{\preserves}[2]{preserves(#1, #2)}

\newcommand{\termk}[3]{\langle #1, #2 \rangle \Downarrow^{#3}}
\newcommand{\term}[2]{\langle #1, #2 \rangle \Downarrow}
\newcommand{\rewrites}[4]{\langle #1, #2 \rangle \longrightarrow \langle #3 , #4 \rangle}
\newcommand{\rewritesbr}[4]{\langle #1, #2 \rangle \longrightarrow \\ &\qquad\langle #3 , #4 \rangle}
\newcommand{\rewritesbrnoi}[4]{\langle #1, #2 \rangle \longrightarrow \\ &\langle #3 , #4 \rangle}
\newcommand{\rewritesbrq}[4]{\langle #1, #2 \rangle \longrightarrow \\ &\quad\langle #3 , #4 \rangle}
\newcommand{\rewritesn}[5]{\langle #1, #2 \rangle \longrightarrow^{#3} \langle #4 , #5 \rangle}
\newcommand{\rewritesstar}[3]{\langle #1, #2 \rangle \longrightarrow^* #3}
\newcommand{\rewritesstarint}[4]{\langle #1, #2 \rangle \longrightarrow^* \langle #3, #4 \rangle}
\newcommand{\rewritesstarbr}[4]{\langle #1, #2 \rangle \longrightarrow^*\\& \langle #3 , #4 \rangle}

\newcommand{\defines}{\stackrel{\text{\normalfont\tiny def}}{=}}
\newcommand{\cvt}[1]{\overline{#1}}
\newcommand{\cvtb}[1]{\underline{#1}}

\newcommand{\ismatch}[2]{\textit{is\_match}(#1, #2)}
\newcommand{\match}[2]{\textit{match}(#1, #2)}
\newcommand{\pvars}[1]{\textit{vars}(#1)}
\newcommand{\resultof}[2]{\textit{eval}(#1, #2)}
\newcommand{\funsof}[1]{\textit{names\_of}(#1)}
\newcommand{\equivext}[3]{\textit{equiv\_ext}(#1, #2, #3)}
\newcommand{\mkcloslist}[1]{\textit{mk\_closlist}(#1)}

\newcommand{\ifclause}{\exc{\textit{error}}{\textit{if\_clause}}{\vtuple{}}}

\maketitle

\section{Introduction}\label{sec:intro}

Most language processors and refactoring tools lack a precise formal specification of how the code is affected by the changes 
they may make. In particular, using a refactoring tool to improve the quality of the code should not change the observable 
behaviour of the program; however, this property is validated only by testing in most cases. Higher assurance can be achieved by 
making formal arguments to verify behaviour preservation, which requires a formal description (e.g., a formal semantics) of 
the programs being refactored, a precise specification how the refactorings affect these programs, and a suitable definition of
program equivalence.

The research presented in this paper is part of a wider project aiming to improve the trustworthiness of Erlang~\cite{cesarini2009erlang}
refactorings via formal verification~\cite{horpacsi2016towards}. As a stepping stone, we formalise Core Erlang~\cite{carlsson2000core}, 
which is a core and intermediate language of Erlang and its compilation process. Beside Erlang, other languages (e.g., Elixir~\cite{elixircomp}) can be translated to Core Erlang, therefore a formalisation of the core language may contribute to the studies of all of the languages in the BEAM family.

In this paper, we are defining a small-step (frame stack style) semantics of Core Erlang extending our previous 
work~\cite{horpacsi2022program} with most of the sequential features of Core Erlang. Based on the formal semantics, 
we define a number of expression equivalence concepts, which we use to prove the correctness of simple Erlang refactorings. All of 
the results presented here, are also formalised with the Coq proof management system~\cite{coreerlang}.

\paragraph{Running example}

We present a simple refactoring in Erlang, which replaces a guard of a function clause with a more effective and readable pattern matching 
(\Cref{fig:refactor1}). Note that in the figure, $f$, $e_1$, $e_2$, and $x$ are metavariables (they can be replaced by concrete 
expressions, atoms, or  variable names) and the side-condition of the refactoring is given as a logical constraint in the \textit{\underline{when}} clause. This example will serve as a running example throughout this paper.

\begin{figure}[ht]
\begin{lstlisting}
         $f$($x$) when length($x$) == 0 -> $e_1$;
         $f$(_) -> $e_2$.
\end{lstlisting}
\vspace{0.2cm}
{\LARGE$\downarrow$} $\normalfont\textit{\underline{when} } x \notin \textit{vars}(e_1)$
\vspace{0.2cm}
\begin{lstlisting}
         $f$([]) -> $e_1$;
         $f$(_) -> $e_2$.
\end{lstlisting}
\caption{A simple function refactoring in Erlang}
\label{fig:refactor1}
\end{figure}

To utilize the formal semantics of Core Erlang presented here, first both of these code chunks are translated to Core Erlang by the 
standard Erlang/OTP compiler (OTP version 24). Next, we encode the Core Erlang programs in the Coq formalisation, and prove their equivalence. In this 
process, we consider the compiler as trusted. In \Cref{fig:code} we show the result of the unoptimised translation of the first function in 
\Cref{fig:refactor1}, after clearing the annotations of Core Erlang~\cite{carlsson2000core}.

\begin{figure}[ht]
\begin{lstlisting}
$f$/1 = fun (_0) ->
  case _0 of
    <$x$> when try let <_1> = call 'erlang':'length'($x$)
                   in call 'erlang':'=='(_1, 0)
             of <Try> -> Try
             catch <T,R> -> 'false'
       -> $e_1$
    <_3> when 'true' -> $e_2$
    <_2> when 'true' ->
      primop 'match_fail'({'function_clause',_2})
  end
\end{lstlisting}
\caption{Core Erlang code of the first function in \Cref{fig:refactor1}}
\label{fig:code}
\end{figure}

\paragraph{Contributions} In this paper, we make the following contributions:

\begin{itemize}
\item A (frame stack) semantics for the sequential subset of Core Erlang including exception handling, which extends and improves on our previous work~\cite{horpacsi2022program}.
\item A definition of termination-based program equivalence concepts (namely: contextual equivalence, equivalence based on logical 
relations, and CIU equivalence).
\item Properties of the program equivalence concepts, and their coincidence.
\item A number of simple (Erlang) expression equivalences (one presented as the running example above).
\item A machine-checked implementation of the results in the Coq proof management system~\cite{coreerlang}.
\end{itemize}

For the proofs of the theorems, lemmas, and examples presented here, we refer to the formalisation~\cite{coreerlang}. 
The rest of the paper is structured as follows. In \Cref{sec:background} we summarise the concepts and our previous work which this 
paper builds on. Thereafter, \Cref{sec:semantics} introduces the formal semantics of sequential Core Erlang. \Cref{sec:equiv} discusses 
program equivalence definitions, followed by a short discussion in \Cref{sec:disc} on the Coq implementation details. \Cref{sec:work} 
briefly discusses related work, while \Cref{sec:conclusion} concludes and points out areas for future work.

\section{Background}\label{sec:background}

In this section, we briefly introduce the concepts of frame stack semantics and program equivalences, and also discuss our 
previous work which we evolve in this paper.

\paragraph{Frame stack semantics}

A frame stack style semantics is a small-step~\cite{plotkin1981structural} operational semantics. It is derived from the reduction-style 
semantics of Wright and Felleisen~\cite{wright1994syntactic}. In the frame stack semantics, the reduction rules can be 
applied in a special reduction context; it is a constructed as a stack of basic evaluation frames. This stack can also be considered
as a continuation of the evaluation. The 
advantage of this style of semantics is that it is simpler to use in a proof assistant since the frame stack is separated into a 
distinct configuration cell (hence it does not need to be inferred like the reduction context in reduction-style semantics).

\paragraph{Program equivalence concepts}

In this paper, we investigate three definitions of program equivalence.
\emph{Contextual equivalence} is a syntactical notion of program equivalence: two expressions are equivalent if 
their behaviours cannot be distinguished in \emph{any syntactical contexts}. Usually, it is burdensome to reason about contextual equivalence 
since it requires induction on the context; however, this notion is needed to express the correctness of local program transformations 
(i.e., equivalent programs can be replaced in arbitrary context without affecting the overall behaviour).

With \emph{equivalence based on logical relations}~\cite{pitts2000operational} two expressions are equivalent when their 
behaviour cannot be distinguished in \emph{equivalent reduction contexts} (i.e., frame stacks). In the frame stack semantics 
reasoning about this equivalence can be carried out by inspecting the semantics of the expressions instead of using induction 
on the contexts.

With \emph{CIU equivalence} (``closed instances of use'' equivalence)~\cite{mason1991equivalence} two expressions are 
equivalent when their behaviour is indistinguishable in \emph{any reduction context}. This notion is the most suitable to reason about 
expressions being equivalent, since it involves only one reduction context (frame stack).

\paragraph{Previous work}

In earlier work~\cite{horpacsi2022program} we have defined frame stack semantics for a distilled, limited variant of Core Erlang, and 
defined the program equivalence concepts mentioned above. Hereby we extend this limited language by adding further 
language elements of Core Erlang: function closures, tuples, maps, sequencing, value lists and value sequences, exceptions and
exception handling. Moreover, we generalise pattern matching, \texttt{let}, \texttt{letrec}, and built-in function call expressions. 
After extending the language and its semantics, we adjust and extend the equivalence relations, corresponding properties and proofs.

Worth mentioning that in prior work we also investigated a big-step style semantics for sequential Core Erlang~\cite{bereczky2020core,bereczky2020machine}, which included studying the semantics of various language elements presented here, allowing us to reuse some of the results achieved there.

\section{Core Erlang Semantics}\label{sec:semantics}

In this section, we discuss the syntax and frame stack semantics of sequential Core Erlang, evaluate an example expression, 
and show a number of semantic properties. For proofs we refer to the machine-checked Coq formalisation~\cite{coreerlang}.

\subsection{Syntax}\label{sec:syntax}

First, we present the syntax of Core Erlang~\cite{carlsson2000core} in \Cref{fig:syntax}. For 
simplicity, we denote lists from the metatheory with $e_1, \dots, e_n$, and non-empty lists with $e_1, e_2, \dots, e_n$. 
We use $x$ to range over variables, $i$ over integers, $a, f$ denote atoms, and $k, n$ are 
used to denote natural numbers. Compared to our previous work~\cite{horpacsi2022program}, here we separate values from expressions, but use 
similar notations for them (e.g., $\etuple{e_1, \dots, e_n}$ is a tuple expression, while $\vtuple{v_1, \dots, v_n}$ is a tuple value).

\begin{figure}[t]
\begin{align*}
p &\in \textit{Pattern} ::= i \hda a \hda x
\hda \econs{p_1}{p_2}
\hda \nil
\hda \etuple{p_1, \dots, p_n} \\
&\hda \emap{\mapitem{p_1^k}{p_1^v}, \dots, \mapitem{p_n^k}{p_n^v}} \\
\textit{ps} &\in \textit{list(Pattern)} ::= \evalues{p_1, \dots, p_n} \\
\textit{cli} &\in \textit{ClosItem} ::= f/k = \fun{x_1, \dots, x_k}{e} \\
\textit{ext} &\in \textit{list(ClosItem)} ::= \textit{cli}_1, \dots, \textit{cli}_n\\
\textit{cl} &\in \textit{Clause} ::= \clause{ps}{e^g}{e^b} \\
v &\in \textit{Val} ::= i \hda a \hda x \hda f/k
\hda \clos{ext}{x_1, \dots, x_n}{e} \\
&\hda \vcons{v_1}{v_2}
\hda \nil
\hda \vtuple{v_1, \dots, v_n} 
\hda \vmap{\mapitem{v_1^k}{v_1^v}, \dots, \mapitem{v_n^k}{v_n^v}} \\
n &\in \textit{NonVal} ::= \fun{x_1, \dots, x_n}{e}
\hda \evalues{e_1, \dots, e_n}
\hda \econs{e_1}{e_2} \\
&\hda \etuple{e_1, \dots, e_n}
\hda \emap{\mapitem{e_1^k}{e_1^v}, \dots, \mapitem{e_n^k}{e_n^v}} \\
&\hda \call{e_m}{e_f}{e_1, \dots, e_n}
\hda \primop{a}{e_1, \dots, e_n} \\
&\hda \apply{e}{e_1, \dots, e_n}
\hda \case{e_1}{\textit{cl}_1; \dots; \textit{cl}_n} \\
&\hda \elet{\varlist{x_1, \dots, x_n}}{e_1}{e_2}
\hda \seq{e_1}{e_2}
\hda \letrec{ext}{e} \\
&\hda \try{e_1}{\varlist{x_1, \dots, x_k}}{e_2}{\varlist{x_{k+1}, \dots, x_{k+n}}}{e_3} \\
e &\in \textit{Exp} ::= n \hda v \\
\textit{vs} &\in \textit{ValSeq} ::= \vvalues{v_1, \dots, v_n} \\
c &\in \textit{ExcClass} ::= \texttt{'throw'} \hda \texttt{'exit'} \hda \texttt{'error'}\\
\textit{exc} &\in \textit{Exception} := \exc{c}{v_r}{v_d} \\
\textit{res} &\in \textit{Result} := \textit{exc} \hda \textit{vs}
\end{align*}
\caption{Syntax of Core Erlang}
\label{fig:syntax}
\end{figure}

The patterns of the language are integers (denoted by numbers), atoms (enclosed in single quotation marks), variables, lists, 
tuples and maps (tilde-enclosed tuples containing key-value pairs denoted with upper indices). The set of values in the language essentially consists of the same constructs, extended with function identifiers ($f/k$, atom-arity pairs) and function closures. Note that we refer to variables and function identifiers as \emph{names} in the rest of the paper.

\paragraph{Closures} Closures are the normal forms of functions (we reuse and adjust their definition from~\cite{bereczky2020machine}). 
Beside a function's parameter list, body expression, they also include a 
list of function definitions ($\textit{ext}$) that can be applied recursively by the body expression (this list is constructed while 
evaluating a \texttt{letrec} expression).

The expressions of the language are either values or non-values, which consist of uncurried function abstractions, 
value lists (denoted by $\evalues{e_1, \dots, e_n}$, usually used in binding expressions), lists, tuples, maps, built-in function (BIF) calls, primitive operations, 
function application, binding expressions (\texttt{let}, \texttt{letrec}, \texttt{case}, \texttt{try}), and sequencing (\texttt{do}).

\paragraph{Evaluation and binding} In Core Erlang all expressions evaluate to either \emph{value sequences} (denoted by 
$\vvalues{v_1, \dots, v_n}$ or $\textit{vs}$) or exceptions (denoted by $\textit{exc}$). Most expressions evaluate to a single value and hence yield singleton value sequences, but value lists ($\evalues{e_1, \dots, e_n}$) evaluate to a value sequence of the same length.

Binding expressions are capable of binding any number of variables (or function identifiers
in case of \texttt{letrec}). For example, if $n$ variables are given in \texttt{try} or \texttt{let} expressions, and $e_1$ evaluates 
to a value sequence of $n$ values, then these values will be bound to the $n$ given variables in $e_2$. This is true for \texttt{case} 
expressions too, but in this case $n$ patterns have to be specified in all of the clauses.

\paragraph{Exceptions} In Erlang implementations~\cite{carlsson2000core}, exceptions (denoted by $\exc{c}{v_r}{v_d}$) consist of an exception class ($c$) and two values describing 
the reason ($v_r$) and additional details ($v_d$) of the exception. These three values are bound inside the \texttt{catch} 
clause of a \texttt{try} expression\footnote{We note that Core Erlang implementations, binding only the first two values in \texttt{catch} 
is also allowed.}.



\subsection{Frame Stacks}\label{sec:framestack}

Next, we define the formal semantics for Core Erlang. We use $\rewrites{K}{r}{K'}{r'}$ to denote reduction steps, where 
the initial configuration consists of the frame stack $K$ and redex $r$, while the final configuration includes the stack $K'$ 
and redex $r'$. Before discussing the rules of the semantics, we define the syntax of redexes, frame stacks, and a number 
of auxiliary definitions.

\begin{figure}
\begin{align*}
r &\in \textit{Redex} ::= \textit{vs}
\hda \textit{exc}
\hda e
\hda \Box \\
%
\textit{id} &\in \textit{FrameId} ::= \textit{tuple} \hda \textit{values} \hda \textit{call}(v_m, v_f) \hda \textit{primop}(a) \hda \textit{map}\\ 
&\hda \textit{app}(v_f) \\
F &\in \textit{Frame} ::= \fparams{\textit{id}}{v_1, \dots, v_{i-1}, \Box, e_{i+1} \dots, e_n}
\hda \econs{e_1}{\Box}
\hda \econs{\Box}{v_2} \\
&\hda \call{\Box}{e_f}{e_1, \dots, e_n}
\hda \call{v_m}{\Box}{e_1, \dots, e_n} \\
&\hda \apply{\Box}{e_1, \dots, e_n}
\hda \case{\Box}{\textit{cl}_1; \dots; \textit{cl}_n} \\
&\hda \case{\textit{vs}}{\clause{ps}{\Box}{e^b} ; \textit{cl}_{2}; \dots; \textit{cl}_n} \\
&\hda \elet{\varlist{x_1, \dots, x_n}}{\Box}{e_2}
\hda \seq{\Box}{e_2} \\
&\hda \try{\Box}{\varlist{x_1, \dots, x_n}}{e_2}{\varlist{x_{k+1}, x_{k+2}, x_{k+3}}}{e_3} \\
K &\in \textit{FrameStack} ::= \idfs \hda F :: K
\end{align*}
\caption{Syntax of redexes, frames, frame stacks}
\label{fig:framesyntax}
\end{figure}

The syntax for frames, frame stacks, and redexes are presented in \Cref{fig:framesyntax}. 
Frames are essentially non-values with one of their subexpression replaced by $\Box$ (they resemble the reduction 
contexts of~\cite{wright1994syntactic}). However, frames do not capture all syntactical contexts. 
Frames capture evaluation order by some of their parameters being values (that have already been evaluated) while others being (non-value) expressions.

Frame stacks are essentially lists: there is the empty stack $\idfs$, and the stack $F :: K$ which denotes the frame stack $K$ with 
the frame $F$ pushed onto it.

\paragraph{Frames for expression lists}

In order to avoid duplicate frames for multiple kinds of expressions containing parameter lists (e.g., tuples, maps, function applications)---which always need to be evaluated in the same way---we introduce frame identifiers, the parameter list frame 
$\fparams{id}{v_1, \dots, v_k, \Box, e_{k+1}, \dots, e_n}$, and the $\Box$ redex to handle empty parameter lists in a uniform way.


\begin{figure*}[t]
\begin{minipage}{\textwidth} 
    \begin{align}
    &\rewrites{K}{\econs{e_1}{e_2}}{\econs{e_1}{\Box} :: K}{e_2} \label{OS:hcons1}\tag{\textsc{SConsTail}}\\[3pt]
    &\rewrites{K}{\elet{\varlist{x_1, \dots, x_n}}{e_1}{e_2}}{\elet{\varlist{x_1, \dots, x_n}}{\Box}{e_2} :: K}{e_1} \label{OS:hlet}\tag{\textsc{SLet}}\\[3pt]
    &\rewrites{K}{\seq{e_1}{e_2}}{\seq{\Box}{e_2} :: K}{e_1} \label{OS:hseq}\tag{\textsc{SSeq}}\\[3pt]
    &\rewrites{K}{\apply{e}{e_1, \dots, e_n}}{\apply{\Box}{e_1, \dots, e_n} :: K}{e} \label{OS:happ}\tag{\textsc{SApp}}\\[3pt]
    &\rewrites{K}{\call{e_m}{e_f}{e_1, \dots, e_n}}{\call{\Box}{e_f}{e_1, \dots, e_n} :: K}{e_m} \label{OS:hcall}\tag{\textsc{SCallMod}}\\[3pt]
    &\rewrites{K}{\primop{a}{e_1, \dots, e_n}}{\fparams{\textit{primop}(a)}{\Box, e_1, \dots, e_n} :: K}{\Box} \label{OS:hprimop}\tag{\textsc{SPrimOp}}\\[3pt]
    &\rewrites{K}{\evalues{e_1, \dots, e_n}}{\fparams{\textit{values}}{\Box, e_1, \dots, e_n} :: K}{\Box}\label{OS:hvals}\tag{\textsc{SVals}}\\[3pt]
    &\rewrites{K}{\etuple{e_1, \dots, e_n}}{\fparams{\textit{tuple}}{\Box, e_1, \dots, e_n} :: K}{\Box}\label{OS:htuple}\tag{\textsc{STuple}}\\[3pt]
    &\rewrites{K}{\emap{\mapitem{e_1^k}{e_1^v}, \mapitem{e_2^k}{e_2^v} \dots, \mapitem{e_n^k}{e_n^v}}}{\fparams{\textit{map}}{\Box, e_1^v, e_2^k, e_2^v, \dots, e_n^k, e_n^v} :: K}{e_1^k}\label{OS:hmap}\tag{\textsc{SMap}}\\[3pt]
    &\rewrites{K}{\case{e}{\textit{cl}_1; \dots; \textit{cl}_n}}{\case{\Box}{\textit{cl}_1; \dots; \textit{cl}_n} :: K}{e}\label{OS:hcase}\tag{\textsc{SCase}}
    \end{align}
\end{minipage}
\caption{Frame stack semantics rules (group \ref{rule:1})}
\label{fig:step1}
\end{figure*}

\subsection{Auxiliary definitions}

For the rest of the paper, we introduce the following concepts:

\begin{itemize}
\item Similarly to our previous work~\cite{horpacsi2022program}, we use $\sub$ to denote capture-avoiding, parallel substitutions. 
Substitutions map names to values. We use $\sub(x)$ to denote the value that is mapped to the name $x$ by the substitution $\sub$.
\item Applying a substitution to a redex (or single value) is denoted by $\subst{r}{\sub}$. If a concrete substitution is given, we use 
$\subst{r}{x_1 \mapsto v_1, \dots, x_n \mapsto v_n}$ which replaces the variables $x_i$ with values $v_i$ in $r$.
\item We also adapt the scoping rules and notations of~\cite{horpacsi2022program} to the extended language. We use $\expscoped{\Gamma}{r}$ 
to denote that the redex (or single value) $r$ contains free names listed in the set $\Gamma$. A redex (or single value) $r$ is closed, if $\expscoped{\emptyset}{r}$.
Moreover, $\subscoped{\Gamma}{\sub}{\Delta}$ denotes that the substitution $\sub$ maps names in $\Gamma$ to values ($v$) such that $\valscoped{\Delta}{v}$.
\item Let $\pvars{p}$ denote the set of variables in pattern $p$.
\item The function $\ismatch{\textit{ps}}{\textit{vs}}$ decides whether the list of patterns $\textit{ps}$ pairwise match to the given 
value sequence $\textit{vs}$. The function $\match{\textit{ps}}{\textit{vs}}$ creates a substitution that includes the result variable-%
value bindings of the successful pattern matching.
\item The function $\funsof{\textit{ext}}$ returns the set of bound function identifiers in the list of function definitions $\textit{ext}$.
\item The function $\mkcloslist{\textit{ext}}$ creates a substitution of closures based on the function definitions in $\textit{ext}$ by 
transforming all function definitions $f/k = \fun{x_1, \dots, x_n}{e}$ of $\textit{ext}$ into $\clos{\textit{ext}}{x_1, \dots, x_n}{e}$ 
($\textit{ext}$ is used in all closures as the collection of recursive function).
For further details we refer to the formalization~\cite{coreerlang}.
\end{itemize}

\subsection{Dynamic Semantics}

In this subsection, we present the rules of the semantics. There are 4 rule categories:

\begin{enumerate}
\item \label{rule:1} Rules that deconstruct an expression by extracting its first redex while putting the rest of the expression in 
the frame stack (\Cref{fig:step1}).
\item \label{rule:2} Rules that modify the top frame of the stack by extracting the next redex and putting back the currently 
evaluated value into this top frame (\Cref{fig:step2}).
\item \label{rule:3} Rules that remove the top frame of the stack and construct the next redex based on this removed frame 
(\Cref{fig:step1}). We also included rules here which immediately reduce an expression without modifying the stack (e.g., \ref{OS:fun}).
\item \label{rule:4} Rules that express concepts of exception creation, handling, or propagation.
\end{enumerate}

\begin{figure*}[t]
\begin{minipage}{\textwidth}
    \begin{align}
    %
    &\rewrites{\econs{e_1}{\Box} :: K}{\vvalues{v_2}}{\econs{\Box}{v_2} :: K}{e_1}\label{OS:hcons2}\tag{\textsc{SConsHead}}\\[3pt]
    &\rewrites{\call{\Box}{e_f}{e_1, \dots, e_n} :: K}{\vvalues{v_m}}{\call{v_m}{\Box}{e_1, \dots, e_n} :: K}{e_f} \label{OS:hcallfun}\tag{\textsc{SCallFun}}\\[3pt]
    &\rewrites{\call{v_m}{\Box}{e_1, \dots, e_n} :: K}{\vvalues{v_f}}{\fparams{\textit{call}(v_m, v_f)}{\Box, e_1, \dots, e_n} :: K}{\Box}\label{OS:hcallparams}\tag{\textsc{SCallParam}}\\[3pt]
    &\rewrites{\apply{\Box}{e_1, \dots, e_n} :: K}{\vvalues{v}}{\fparams{\textit{apply}(v)}{\Box, e_1, \dots, e_n} :: K}{\Box}\label{OS:happarams}\tag{\textsc{SAppParam}}\\[3pt]
    %
    &\rewrites{\case{\Box}{\clause{\textit{ps}}{e^g}{e^b}; \textit{cl}_2; \dots; \textit{cl}_n} :: K}{\textit{vs}}{\case{\Box}{\textit{cl}_2; \dots; \textit{cl}_n} ::K}{\textit{vs}} \hspace{2em}(\text{if } \neg\ismatch{ps}{vs}) \label{OS:casefail}\tag{\textsc{SCaseFail}}\\[3pt]    
    \begin{split}    
    &\rewritesbr{\case{\Box}{\clause{\textit{ps}}{e^g}{e^b}; \textit{cl}_2; \dots; \textit{cl}_n} :: K}{\textit{vs}}{\case{\textit{vs}}{\clause{\textit{ps}}{\Box}{\subst{e^b}{\match{\textit{ps}}{\textit{vs}}}}; \textit{cl}_2; \dots; \textit{cl}_n} :: K}{\subst{e^g}{\match{\textit{ps}}{\textit{vs}}}}\qquad (\text{if } \ismatch{ps}{vs})
    \end{split} \label{OS:casesuccess}\tag{\textsc{SCaseSuccess}}\\[3pt]
    &\rewrites{\case{\textit{vs}}{\clause{\textit{ps}}{\Box}{e^b}; \textit{cl}_2; \dots; \textit{cl}_n} :: K}{\vvalues{\texttt{'false'}}}{\case{\Box}{\textit{cl}_2; \dots; \textit{cl}_n} :: K}{\textit{vs}}\label{OS:casefalse}\tag{\textsc{SCaseFalse}}\\[3pt]
    %
    &\rewrites{\fparams{\textit{id}}{\Box, e_1, e_2, \dots, e_n} :: K}{\Box}{\fparams{\textit{id}}{\Box, e_2, \dots, e_n} :: K}{e_1}\label{OS:params0}\tag{\textsc{SParams$_0$}} \qquad(\text{if } \textit{id} \neq \textit{map})\\[3pt]    
    &\rewrites{\fparams{\textit{id}}{v_1, \dots, v_{i-1}, \Box, e_{i+1}, e_{i+2}, \dots, e_n} :: K}{\vvalues{v_i}}{\fparams{\textit{id}}{v_1, \dots, v_{i-1}, v_i, \Box, e_{i + 2}, \dots, e_n} :: K}{e_{i+1}}\label{OS:params}\tag{\textsc{SParams}}
    \end{align}
\end{minipage}
    \caption{Frame stack semantics rules (group \ref{rule:2})}
\label{fig:step2}
\end{figure*}

Next, we discuss a number of the more complex rules, starting with rules of group~\ref{rule:1} (\Cref{fig:step1}):

\begin{itemize}
\item \ref{OS:hprimop}, \ref{OS:htuple}, and \ref{OS:hvals} reduce expressions with parameter lists. They put a parameter list frame 
(with their corresponding frame identifier) on the top of the frame stack. To avoid handling empty parameter lists separately for 
each language element, $\Box$ is put into the final configuration, which will be handled by \ref{OS:cparams0} in case of an empty, or \ref{OS:params0} in case of a non-empty parameter list.
\item \ref{OS:hmap} starts the evaluation of a non-empty map expression by creating a parameter list frame. In this case, the use of 
$\Box$ can be avoided, since there is at least one key expression. Note that empty maps are handled separately to satisfy that 
the sum of subexpressions and values in parameter list frames for maps is always an odd number (we refer to the description of \ref{OS:cparams} for more insights).
\item The rest of the rules of \Cref{fig:step1} extract the first redex of the given expression, and push the remaining parts onto the stack. We note that lists are evaluated in a right-to-left order in Core Erlang, this is why $e_2$ is extracted 
first in \ref{OS:hcons1}.
\end{itemize}

Now we draw attention to the rules in groups~\ref{rule:2} (\Cref{fig:step2}) and~\ref{rule:3} (\Cref{fig:step3}). Observe that all of these expect the redex in the initial configuration to be a singleton value sequence, except for the binding expressions (in line with that we said in \Cref{sec:syntax}), single-step reduction rules, and technical rules involving $\Box$.

\begin{itemize}
\item \ref{OS:params0} starts the evaluation of non-empty parameter lists. This is one of the two rules that expects $\Box$ in the 
initial configuration. If there are some expressions in the parameter list frame on the top of the stack, this rule extracts the first 
one. Note that this rule cannot be used for map frames.
\item \ref{OS:params} extracts the next redex ($e_{i+1}$) from the parameter list frame on the top of the stack, if the $i$th 
expression has already been reduced to a singleton value sequence. The item in this singleton sequence is put back into the frame.
\item \ref{OS:hcallparams}, \ref{OS:happarams} express reductions for frames with parameter lists, and behave the same way as 
described above for \ref{OS:hprimop}, \ref{OS:htuple}, and \ref{OS:hvals}.
\item \ref{OS:casefail} expresses the evaluation of a \texttt{case} expression if the pattern matching failed. 
In this case, the first clause can be removed, and the next clause needs to be checked.
\item \ref{OS:casesuccess} expresses the evaluation of a \texttt{case} expression if the pattern matching succeeds. In this case, the 
next redex to evaluate is the guard expression of the current clause, substituted by the result of the pattern matching. Note that the 
substitution is also applied to the body expression of the current clause.
\item \ref{OS:casefalse} expresses when a guard of a clause evaluates to \texttt{'false'}. In this case, the first clause can be removed,
and the next clause needs to be checked.
\item The rest of the rules of \Cref{fig:step2} extract the next redex from the top frame of the stack, and put back the result value 
(which is inside a singleton value sequence) into frame.
\end{itemize}

\begin{figure*}[t]
\begin{minipage}{\textwidth}
    \begin{align}
    &\rewrites{K}{\emap{}}{K}{\vvalues{\vmap{}}}\label{OS:mapzero}\tag{\textsc{PMap$_0$}}\\[3pt]
    &\rewrites{K}{\fun{x_1, \dots, x_n}{e}}{K}{\vvalues{\clos{\emptyset}{x_1, \dots, x_n}{e}}}\label{OS:fun}\tag{\textsc{PFun}}\\[3pt]
    &\rewrites{K}{\letrec{\textit{ext}}{e}}{K}{\subst{e}{\mkcloslist(\textit{ext})}}\label{OS:hletrec}\tag{\textsc{PLetRec}}\\[3pt]
    &\rewrites{K}{v}{K}{\vvalues{v}}\label{OS:value}\tag{\textsc{PValue}}\\[3pt]
    &\rewrites{\fparams{\textit{id}}{v_1, \dots, v_n, \Box} :: K}{\Box}{K}{\resultof{id}{v_1, \dots, v_n}} \qquad(\text{if } \textit{id} \neq \textit{map}) \label{OS:cparams0}\tag{\textsc{PParams$_0$}}\\[3pt]
    &\rewrites{\fparams{\textit{id}}{v_1, \dots, v_{n-1}, \Box} :: K}{\vvalues{v_n}}{K}{\resultof{id}{v_1, \dots, v_n}}\label{OS:cparams}\tag{\textsc{PParams}}\\[3pt]
    %
    &\rewrites{\econs{\Box}{v_2} ::K}{\vvalues{v_1}}{K}{\vvalues{\vcons{v_1}{v_2}}}\label{OS:ccons}\tag{\textsc{PCons}}\\[3pt]    
    &\rewrites{\case{\textit{vs}}{\clause{\textit{ps}}{\Box}{e^b}; \textit{cl}_2; \dots; \textit{cl}_n} :: K}{\vvalues{\texttt{'true'}}}{K}{e^b}\label{OS:casetrue}\tag{\textsc{PCaseTrue}}\\[3pt]
    &\rewrites{\elet{\varlist{x_1, \dots, x_n}}{\Box}{e_2} :: K}{\vvalues{v_1, \dots, v_n}}{K}{\subst{e_2}{x_1 \mapsto v_1, \dots, x_n \mapsto v_n}}\label{OS:clet}\tag{\textsc{PLet}}\\[3pt]
    &\rewrites{\seq{\Box}{e_2} :: K}{\vvalues{v_1}}{K}{e_2}\label{OS:cseq}\tag{\textsc{PSeq}}
    \end{align}
\end{minipage}
    \caption{Frame stack semantics rules (group \ref{rule:3})}
\label{fig:step3}
\end{figure*}

We use the auxiliary function $\resultof{\textit{id}}{v_1, \dots, v_n}$ that constructs a redex based on a frame identifier and a 
parameter list. We provide an informal overview of its definition here, and for the precise definition, we refer to the 
formalisation~\cite{coreerlang}. If
\begin{itemize}
\item $\textit{id} = \textit{app}(v_f)$ and $v_f = \clos{ext}{x_1, \dots, x_n}{e}$, then \\
$\resultof{\textit{app}(v_f)}{v_1, \dots, v_n} =\subst{e}{\mkcloslist{\textit{ext}}, x_1 \mapsto v_1, \dots,$ $x_n$ $\mapsto v_n}$
\item $\textit{id} = \textit{app}(v_f)$ and $v_f$ is not a closure, or has an incorrect number of formal parameters, the result is an exception.
\item $\textit{id} = \textit{tuple}$, then $\resultof{\textit{tuple}}{v_1, \dots, v_n} = \vtuple{v_1, \dots, v_n}$.
\item $\textit{id} = \textit{values}$, then $\resultof{\textit{values}}{v_1, \dots, v_n} = \vvalues{v_1, \dots, v_n}$.
\item $\textit{id} = \textit{map}$ and $n$ is an even number, then
\begin{gather*}
\qquad\resultof{\textit{map}}{v_1, \dots, v_n} = \vmap{\mapitem{v_1}{v_2}, \dots, \mapitem{v_{k-1}}{v_k}},
\end{gather*}
where the $k \le n$ result values inside the map are obtained by eliminating duplicate keys and their associated values. 
\item $\textit{id} = \textit{call}(a_m, a_f)$, then $\resultof{\textit{call}(a_m, a_f)}{v_1, \dots, v_n}$ simulates the behaviour of 
the built-in functions of (Core) Erlang.
\item $\textit{id} = \textit{primop(a)}$, then $\resultof{\textit{primop(a)}}{v_1, \dots, v_n}$ simulates the behaviour of primitive operations in Core Erlang.
\end{itemize}

Thereafter, we highlight some rules from group~\ref{rule:3} (\Cref{fig:step3}):

\begin{itemize}
\item \ref{OS:mapzero}, \ref{OS:fun}, and \ref{OS:hletrec} express single step reductions that do not include the manipulation of the frame 
stack.
\item \ref{OS:value} reduces a value to a singleton value sequence. In most cases, this rule is used to evaluate atoms, integers, 
and empty lists, since these values do not have a corresponding expression, in contrast to tuples, maps, and non-empty lists.
\item \ref{OS:cparams0} handles the evaluation of empty parameter lists (note that maps are handled separately with \ref{OS:mapzero}). This is the other rule (besides \ref{OS:params0}) that expects $\Box$ in the initial configuration as the redex.
\item \ref{OS:cparams} handles parameter lists. If all of the expressions have been evaluated to values, then based on the frame identifier 
the next redex is constructed with $\resultof{\textit{id}}{v_1, \dots, v_n}$. Note that if the frame identifier was $\textit{map}$, then 
$n$ is required to be an even number (i.e., there is an odd number of subvalues in the top frame, and the last value is in the second 
configuration cell).
\item \ref{OS:casetrue} is used when the guard expression of the clause is evaluated to \texttt{'true'}. The next redex is the body 
expression of the same clause. Note that the bindings obtained from the successful pattern matching are already substituted by \ref{OS:casesuccess}.
\end{itemize}

\begin{figure*}[t]
\begin{minipage}{\textwidth}
    \begin{align}
    &\rewrites{\case{\Box}{\emptyset}::K}{\textit{vs}}{K}{\ifclause} \label{OS:ifclause}\tag{\textsc{ExcCase}}\\[3pt]
    &\rewrites{K}{\try{e_1}{\varlist{x_1, \dots, x_n}}{e_2}{\varlist{x_{k+1}, \dots, x_{k+n}}}{e_3}}{\try{\Box}{\varlist{x_1, \dots, x_n}}{e_2}{\varlist{x_{k+1}, \dots, x_{k+n}}}{e_3} :: K}{e_1}\label{OS:htry}\tag{\textsc{STry}}\\[3pt]
    &\rewrites{\try{\Box}{\varlist{x_1, \dots, x_n}}{e_2}{\varlist{x_{k+1}, \dots, x_{k+n}}}{e_3} :: K}{\vvalues{v_1, \dots, v_n}}{K}{\subst{e_2}{x_1 \mapsto v_1, \dots, x_n \mapsto v_n}}\label{OS:trynormal}\tag{\textsc{PTry}}\\[3pt]  
    &\rewrites{\try{\Box}{\varlist{x_1, \dots, x_n}}{e_2}{\varlist{x_{k+1}, \dots, x_{k+3}}}{e_3} :: K}{\exc{c}{v_r}{v_d}}{K}{\subst{e_3}{x_{k+1}\mapsto c, x_{k+2} \mapsto v_r, x_{k+3} \mapsto v_d }}\label{OS:tryexc}\tag{\textsc{ExcTry}}\\[3pt]
    &\rewrites{F :: K}{\exc{c}{v_r}{v_d}}{K}{\exc{c}{v_r}{v_d}} \qquad (\text{if } \textit{F} \neq \try{\Box}{\varlist{x_1, \dots, x_n}}{e_2}{\varlist{x_{k+1}, \dots, x_{k + n}}}{e_3})\label{OS:excprop}\tag{\textsc{ExcProp}}
    \end{align}
\end{minipage}
    \caption{Frame stack semantics rules (group \ref{rule:4})}
\label{fig:step4}
\end{figure*}

Finally, we explain the rules for exception creation, handling and propagation (\Cref{fig:step4}):

\begin{itemize}
\item \ref{OS:ifclause} is used when none of the clauses of a \texttt{case} expression matched, or all of the guards of the 
matching clauses evaluated to \texttt{'false'}. In these cases an exception is raised. Note that this is not the only option 
to raise exceptions: exceptions can be the result of computing $\resultof{\textit{id}}{v_1, \dots, v_n}$.
\item \ref{OS:htry} extracts the first redex from a \texttt{try} expression for evaluation. (This rule could also belong to group~\ref{rule:1}.)
\item \ref{OS:trynormal} is used when the first subexpression of a \texttt{try} expression evaluated to a value sequence. In this 
case (if the number of variables are correct) the execution continues with the expression of the first clause substituted with 
the resulting variable-value bindings.
\item \ref{OS:tryexc} is used when the first subexpression of a \texttt{try} expression evaluated to an exception. In this case, three 
variables are bound to the parts of the exceptions in the expression of the \texttt{catch} clause, and the evaluation continues with 
this redex.
\item \ref{OS:excprop} describes exception propagation. If the first frame is not an exception handler, it is removed from the stack.
\end{itemize}

\paragraph{The evaluation relation.}
 
Now we can define the step-indexed, reflexive, transitive closure of the reductions as usual (denoted by $\rewritesn{K}{r}{n}{K'}{r'}$ when the number of reduction steps is relevant, $\rewritesstarint{K}{r}{K'}{r'}$ when it is not).

According to~\cite{horpacsi2022program,pitts2000operational} (and \Cref{thm:ciuterm}) it is sufficient to reason about termination 
for programs to be equivalent, thus next we define termination.
A redex terminates in frame stack $K$ if it can be evaluated either to a value sequence or exception.

\begin{definition}[Termination]
\begin{align*}
\termk{K}{r}{n} &:= \exists \textit{res}: \rewritesn{K}{r}{n}{\idfs}{\textit{res}}\\
\term{K}{r} &:= \exists n: \termk{K}{r}{n}
\end{align*}
\end{definition}

Finally, we highlight some properties of the semantics which were heavily used in the proofs on program equivalence. The first property 
expresses that adding frames to the bottom of the stack (denoted by $\concat$) does not affect the behaviour.

\begin{theorem}[Extend frame stack]\label{thm:extendframes}
For all frame stacks $K_1$, $K_2$, $K'$, redexes $r_1$, $r_2$, and step counters $n$, if $\rewritesn{K_1}{r_1}{n}{K_2}{r_2}$, then 
$\rewritesn{K_1 \concat K'}{r_1}{n}{K_2 \concat K'}{r_2}$.
\end{theorem}

The next property expresses whenever a redex terminates in a frame stack, the redex can be evaluated to a value sequence or exception in 
the empty frame stack.

\begin{theorem}[Termination and reductions]
For all frame stacks $K$, redexes $r$, and step counters $n$, if $\termk{K}{r}{n}$ then $\exists \textit{res}, k \le n: \rewritesn{\idfs}{r}{k}{\idfs}{\textit{res}}$.
\end{theorem}

The next two properties show that the frame stack can be merged into the evaluable expression. We use $\fsubst{F}{e}$ to substitute an 
expression $e$ into the $\Box$ of frame $F$. While this operation is a syntactical replacement for most frames, there is one exception: 
\[\case{\textit{vs}}{\clause{\textit{ps}}{\Box}{e^b}; \textit{cl}_2; \dots; \textit{cl}_n}\]

If this frame is on the top of the stack, the semantics has already substituted 
the pattern variables of \textit{ps}, thus these variables should not be substituted again in the expression that replaces $\Box$ (and in $e^b$ too). 
Thus for this case we define the substitution in the following way:
\begin{align*}
&\fsubst{(\case{\textit{vs}}{\clause{\textit{ps}}{\Box}{e^b}; \textit{cl}_2; \dots; \textit{cl}_n})}{e^g} := \\
&\quad\case{\vvalues{}}{\\
&\qquad\clause{\evalues{}}{e^g}{e^b}; \\
&\qquad\clause{\evalues{}}{\texttt{'true'}}{\case{\textit{vs}}{\textit{cl}_2; \dots; \textit{cl}_n}}\\&\quad}
\end{align*}
With this definition, we highlight the following two properties of the frame stack. Note that a frame is closed if substituted by a 
closed expression, the result is closed, and a frame stack is closed, when all of its frames are closed.

\begin{theorem}[Remove frame]\label{thm:putback}
For all closed frames $F$, closed expressions $e$, and all frame stacks $K$, if $\term{F :: K}{e}$ then $\term{K}{\fsubst{F}{e}}$.
\end{theorem}

The next theorem is the opposite of the previous one, allowing a context frame to be pushed to the stack.

\begin{theorem}[Add frame]\label{thm:putbackrev}
For all closed frames $F$, closed expressions $e$, and all frame stacks $K$, if $\term{K}{\fsubst{F}{e}}$ then $\term{F :: K}{e}$.
\end{theorem}

\subsection{Example}

Next, we show an example on using the frame stack semantics. We recall the expression from \Cref{fig:code}, and replace the metavariables 
with concrete values:

\smallskip
\begin{lstlisting}
'f'/1 = fun (_0) ->
  case _0 of
    <L> when try let <_1> = call 'erlang':'length'(L)
                   in call 'erlang':'=='(_1, 0)
             of <Try> -> Try
             catch <T,R> -> 'false'
       -> 1
    <_3> when 'true' -> 2
    <_2> when 'true' ->
      primop 'match_fail'({'function_clause',_2})
  end
\end{lstlisting}
\smallskip

We note that the semantics requires the \texttt{catch} clauses to bind three variables (which was based on the language 
specification~\cite{carlsson2000core}), while the compiler also works with only two, thus in the actual formalisation, we introduced a third variable. 

We denote the clauses of the \texttt{case} subexpression with $\textit{cl}_1$, $\textit{cl}_2$, $\textit{cl}_3$, the \texttt{try} 
subexpression with \textit{try}, and the \texttt{let} subexpression with \textit{let}. Suppose that we apply this function to a value $v$. 
In the first three steps the singleton value $v$ is evaluated from the head of the \texttt{case} expression. This involves transforming 
$v$ into a singleton value sequence in the second step.
In the next steps, rules from group \ref{rule:1} are used to deconstruct the complex expression until the \texttt{call} expression is reached.

\begin{align*}
&\rewritesbrnoi{\idfs}{\case{v}{\textit{cl}_1; \textit{cl}_2; \textit{cl}_3}}
{\case{\Box}{\textit{cl}_1; \textit{cl}_2; \textit{cl}_3} :: \idfs}{v}\longrightarrow\\
&\rewritesbrnoi{\case{\Box}{\textit{cl}_1; \textit{cl}_2; \textit{cl}_3} :: \idfs}{\vvalues{v}}
{\case{\vvalues{v}}{\clause{\varlist{\texttt{L}}}{\Box}{\texttt{1}}; \textit{cl}_2; \textit{cl}_3} :: \idfs}{\textit{try}}\longrightarrow\\
&\rewritesbrnoi{\try{\Box}{\varlist{\texttt{Try}}}{\texttt{Try}}{\varlist{\texttt{T}, \texttt{R}}}{\texttt{'false'}} :: \\
  &\qquad\case{\vvalues{v}}{\clause{\varlist{\texttt{L}}}{\Box}{\texttt{1}}; \textit{cl}_2; \textit{cl}_3} :: \idfs}{\textit{let}}
{\elet{\varlist{\texttt{\_1}}}{\Box}{\call{\texttt{'erlang'}}{\texttt{'=='}}{\texttt{\_1}, \texttt{0}}}::\\
  &\qquad\try{\Box}{\varlist{\texttt{Try}}}{\texttt{Try}}{\varlist{\texttt{T}, \texttt{R}}}{\texttt{'false'}} :: \\
  &\qquad\case{\vvalues{v}}{\clause{\varlist{\texttt{L}}}{\Box}{\texttt{1}}; \textit{cl}_2; \textit{cl}_3} :: \idfs}{\\
  &\qquad\call{\texttt{'erlang'}}{\texttt{'length'}}{v}}
\end{align*}

For readability, we show the evaluation of this \texttt{call} expression separately, and denote the current stack with $K$. Note that at this 
point the variable \texttt{L} has already been replaced by $v$. First, the module and then the function expression is turned into a singleton 
value sequence and put back into the frame stack (we merged these steps below). Then, the parameters are evaluated using a parameter list 
frame. In this case, there is one parameter, thus first \ref{OS:params0} is used, then $v$ is reduced to a singleton value sequence, and 
finally, the use of \ref{OS:cparams} concludes these reduction steps.

\begin{align*}
&\rewritesstarbr{K}{\call{\texttt{'erlang'}}{\texttt{'length'}}{v}}
{\call{\Box}{\texttt{'length'}}{v} :: K}{\texttt{'erlang'}}\longrightarrow^*\\
&\rewritesstarbr{\call{\texttt{'erlang'}}{\Box}{v} :: K}{\texttt{'length'}}
{\fparams{\textit{call}(\texttt{'erlang'},\texttt{'length'})}{\Box, v} :: K}{\Box}\longrightarrow\\
&\rewritesbrnoi{\fparams{\textit{call}(\texttt{'erlang'},\texttt{'length'})}{\Box} :: K}{v}
{\fparams{\textit{call}(\texttt{'erlang'},\texttt{'length'})}{\Box} :: K}{\vvalues{v}}\longrightarrow\\
&\langle K , \resultof{\textit{call}(\texttt{'erlang'},\texttt{'length'})}{v} \rangle \label{ex:steplet}\tag{\textsc{Result}}
\end{align*}

At this point, the result depends on the value $v$. First, let us suppose that $v = \nil$, and proceed with the evaluation. In this case, 
the result of calling \texttt{'length'} is \texttt{0}. Let us denote the current frame stack without the first \texttt{let} frame with $K_1$, 
and the stack we get by removing the \texttt{try} frame from $K_1$ with $K_2$. The next step is to evaluate the equality check (\texttt{'=='}) 
expression inside \texttt{let}, which is done analogously to calling \texttt{length} above. The result is \texttt{'true'}, which is not an exception, thus it is propagated through the \texttt{try} expression. This means that the guard is true of the \texttt{case} expression, thus 
\ref{OS:casetrue} is used followed by reducing \texttt{1} into a singleton value sequence.

\begin{align*}
&\rewritesbrnoi{\elet{\varlist{\texttt{\_1}}}{\Box}{\call{\texttt{'erlang'}}{\texttt{'=='}}{\texttt{\_1}, \texttt{0}}} :: K_1}{\vvalues{\texttt{0}}}
{K_1}{\call{\texttt{'erlang'}}{\texttt{'=='}}{\texttt{0}, \texttt{0}}}\longrightarrow^*\\
&\rewritesstarbr{\try{\Box}{\varlist{\texttt{Try}}}{\texttt{Try}\\&\qquad}{\varlist{\texttt{T}, \texttt{R}}}{\texttt{'false'}} :: K_2}{\vvalues{\texttt{'true'}}}
{\case{\vvalues{\nil}}{\clause{\varlist{\texttt{L}}}{\Box}{\texttt{1}}; \textit{cl}_2; \textit{cl}_3} :: \idfs}{\vvalues{\texttt{'true'}}}\longrightarrow^*\\
&\langle \idfs, \vvalues{\texttt{1}} \rangle
\end{align*}

Next, we discuss the evaluation for another value. Suppose that $v = \texttt{0}$ when the evaluation reached the point in 
equation~\ref{ex:steplet}. In this case, the result of calling \texttt{'length'} is a bad argument exception (we denote it with 
$\textit{badarg}$). In this case, the next reduction with \ref{OS:excprop} removes the frame for \texttt{let}, then the 
exception is handled by the frame for \texttt{try} with \ref{OS:tryexc}. 
The expression in the \texttt{catch} clause is \texttt{'false'}, thus the next clause of the \texttt{case} expression 
is checked (\ref{OS:casefalse}). In this clause both the pattern matching succeeds, and the guard evaluates to \texttt{'true'}, thus the 
final result is $\vvalues{2}$ in this case.

\begin{align*}
&\rewritesbrnoi{\elet{\varlist{\texttt{\_1}}}{\Box}{\call{\texttt{'erlang'}}{\texttt{'=='}}{\texttt{\_1}, \texttt{0}}} :: K_1}{\textit{badarg}}
{\try{\Box}{\varlist{\texttt{Try}}}{\texttt{Try}\\&\qquad}{\varlist{\texttt{T}, \texttt{R}}}{\texttt{'false'}} :: K_2}{\textit{badarg}}\longrightarrow\\
&\rewritesbrnoi{\case{\vvalues{\texttt{0}}}{\clause{\varlist{\texttt{L}}}{\Box}{\texttt{1}}; \textit{cl}_2; \textit{cl}_3} :: \idfs}{\texttt{'false'}}
{\case{\vvalues{\texttt{0}}}{\clause{\varlist{\texttt{L}}}{\Box}{\texttt{1}}; \textit{cl}_2; \textit{cl}_3} :: \idfs}{\vvalues{\texttt{'false'}}}\longrightarrow\\
&\rewritesstarbr{\case{\Box}{\clause{\evalues{\texttt{\_3}}}{\texttt{'true'}}{\texttt{2}}; \textit{cl}_3} :: \idfs}{\texttt{0}}
{\case{\vvalues{\texttt{0}}}{\clause{\evalues{\texttt{\_3}}}{\Box}{\texttt{2}}; \textit{cl}_3} :: \idfs}{\texttt{'true'}}\longrightarrow^*\\
&\langle \idfs, \vvalues{\texttt{2}} \rangle
\end{align*}

For more details and examples, we refer to the formalisation~\cite{coreerlang}.

\section{Program Equivalence}\label{sec:equiv}

In this section, we show three concepts of program equivalence we investigated and formalised based on our previous 
work~\cite{horpacsi2022program}; program equivalence based on logical relations, CIU equivalence, and contextual equivalence.

\subsection{Program Equivalence Based on Logical Relations}\label{sec:logrel}

First, we define program equivalence with logical relations based on the techniques of 
Pitts~\cite{pitts1997operationally,pitts2000operational}. Since Core Erlang is a dynamically typed language, we cannot rely on 
types (a typing judgement) to express the mutual definitions, so we formalise the relations using 
step-indexing~\cite{ahmed2006stepindexed} (following the approach of Wand et al.~\cite{wand2018contextual}). We start by defining 
the relations for closed expressions, values, frame stacks, and exceptions.

\begin{definition}[Logical relations for closed expressions, values, exceptions and frame stacks]\label{def:logrelclosed}
First, we define the logical relation for expressions. We denote the set of related expressions with $\mathbb{E}_n$, where $n$ is a 
step counter. Two expressions are related, when the first one terminates in at most $n$ steps in a frame stack, the second also terminates 
(in any number of steps) in all frame stacks that are related to the stack in the first termination.

Note that this first definition is not (yet) about redexes, only about expressions. This decision correlates to the 
definition of (syntactical) contextual equivalence in \Cref{sec:ctx}, and this definition will coincide with contextual equivalence.
%
\begin{align*}
    &(e_1, e_2) \in \mathbb{E}_n \defines
    (\forall m \le n, K_1, K_2: (K_1, K_2) \in \mathbb{K}_m \implies\\ 
    &\qquad\termk{K_1}{e_1}{m} \implies \term{K_2}{e_2})
\end{align*}
We denote the set of related frame stacks $\mathbb{K}_n$, where $n$ is a step counter. 
Two stacks are related whenever the first one terminates in at most $n$ steps in a configuration with a value sequence, exception, or $\Box$, 
then the second stack also terminates (in any number of steps) in all configurations which contain value sequence, exception, or $\Box$ that 
are related to the value sequence, exception, or $\Box$ in the other configuration.
\begin{align*}
    &(K_1, K_2) \in \mathbb{K}_n \defines\\
    &(\forall m \le n, v_1, v_1', \dots, v_n, v_n': (v_1, v_1'), \dots, (v_n, v_n') \in \mathbb{V}_m \implies \\
    &\qquad\termk{K_1}{\vvalues{v_1, \dots, v_n}}{m} \implies \term{K_2}{\vvalues{v_1', \dots, v_n'}}) \land \\
    &(\forall m \le n, \textit{exc}_1, \textit{exc}_2: (\textit{exc}_1, \textit{exc}_2) \in \mathbb{X}_m \implies \\
    &\qquad\termk{K_1}{\textit{exc}_1}{m} \implies \term{K_2}{\textit{exc}_2}) \land \\
    &(\forall m \le n: \termk{K_1}{\Box}{m} \implies \term{K_2}{\Box})
\end{align*}
Next, we define the concept of related values (their set is denoted by $\mathbb{V}_n$, where $n$ is a step counter). 
This relation defines the base cases of the mutual definitions. Two atoms, integers are related when they are equal. 
Two empty lists are always related, while non-empty value lists are related when their subvalues are related.  Similarly, 
tuples and maps are related when they are related element-wise.
Two closures are related, if their bodies---substituted with their recursive function definitions (\textit{ext}, \textit{ext'}) 
and pairwise-related actual parameters---are related expressions. In this case, we do not require the recursive definitions to be 
related, only the termination of the body expression matters. We also highlight the $<$ relation on the step counters in this relation 
to ensure well-formed recursion.
%
\begin{align*}
    &(i_1, i_2) \in \mathbb{V}_n \defines i_1 = i_2\qquad\qquad
    (a_1, a_2) \in \mathbb{V}_n \defines a_1 = a_2\\
    &(\nil, \nil) \in \mathbb{V}_n \defines \textit{true}\\
    &(\vcons{v_1}{v_2}, \vcons{v_1'}{v_2'}) \in \mathbb{V}_n \defines (v_1, v_1'), (v_2, v_2') \in  \mathbb{V}_n\\
    &(\vtuple{v_1, \dots, v_n}, \vtuple{v_1', \dots, v_n'}) \in \mathbb{V}_n \defines (v_1, v_1'), \dots, (v_n, v_n') \in  \mathbb{V}_n\\
    &(\vmap{\mapitem{v_1^k}{v_1^v}, \dots, \mapitem{v_n^k}{v_n^v}}, \vmap{\mapitem{v_1^{k'}}{v_1^{v'}}, \dots, \mapitem{v_n^{k'}}{v_n^{v'}}}) \in \mathbb{V}_n\\
    &\quad\defines(v_1^k, v_1^{k'}), (v_1^v, v_1^{v'}), \dots, (v_n^k, v_n^{k'}), (v_n^v, v_n^{v'}) \in  \mathbb{V}_n\\
    &(\clos{ext}{x_1, \dots, x_n}{e}, \clos{ext'}{x_1, \dots, x_n}{e'}) \in \mathbb{V}_n \defines \\
    &\qquad(\forall m < n: \forall v_1, v_1', \dots, v_n, v_n': (v_1, v_1'), \dots,  (v_n, v_n') \in \mathbb{V}_m \implies\\
    &\qquad\qquad(\subst{e}{\mkcloslist{\textit{ext}}, x_1 \mapsto v_1, \dots, x_n \mapsto v_n},\\
    &\qquad\qquad\subst{e'}{\mkcloslist{\textit{ext}'}, x_1 \mapsto v_1', \dots, x_n \mapsto v_n'}) \in \mathbb{E}_m)
\end{align*}
Finally, we define the logical relation for exceptions (denoted by $\mathbb{X}_n$, where $n$ is a step counter). 
Two exceptions are related, when their three subvalues are pairwise related (note that the exception classes are always atoms, thus 
they are related if they are equal).
\begin{align*}
&(\exc{c}{v_r}{v_d}, \exc{c'}{v_r'}{v_d'}) \in \mathbb{X}_n \defines\\ 
&\qquad c = c' \land (v_r, v_r') \in \mathbb{V}_n \land (v_d, v_d') \in \mathbb{V}_n
\end{align*}
\end{definition}

Next, we also define logical relations for redexes (i.e., not only for expressions but also for values, exceptions and holes). This concept coincides with CIU equivalence (\Cref{sec:ciu}).

\begin{definition}[Logical relation for redexes]
\begin{align*}
    &(r_1, r_2) \in \mathbb{R}_n \defines
    (\forall m \le n, K_1, K_2: (K_1, K_2) \in \mathbb{K}_m \implies\\ 
    &\qquad\termk{K_1}{r_1}{m} \implies \term{K_2}{r_2})
\end{align*}
\end{definition}

Similarly to the related work, relations with higher indices can distinguish more expressions, frame stacks, values, exceptions, and redexes.

\begin{theorem}[Monotonicity of the logical relations]
For all step counters $n$, $m$, if $m \le n$, then $R_n \subseteq R_m$ for $R \in \{\mathbb{E}, \mathbb{K}, \mathbb{V}, \mathbb{X}, \mathbb{R}\}$.
\end{theorem}

Next, we generalise the relations for closed elements of the syntax to open elements too. For this, first we define 
related substitutions.

\begin{definition}[Logical relations with closing substitutions]\label{def:logrelopen}
We denote the set of related substitutions with $\mathbb{G}^\Gamma_n$, where $n$ is the usual step counter,
and $\Gamma$ is the set of free variables that are substituted with closed values by the substitutions.
\begin{align*}
(\sub_1, \sub_2) \in \mathbb{G}^\Gamma_n \defines&\ \subscoped{\Gamma}{\sub_1}{\emptyset} \land \subscoped{\Gamma}{\sub_2}{\emptyset} \ \land\\
&(\forall x \in \Gamma: (\sub_1(x), \sub_2(x)) \in \mathbb{V}_n)
\end{align*}
With the concept of related closing substitutions, we can define the logical relations for open expressions, values, exceptions, redexes.
\begin{align*}
(v_1, v_2) \in \mathbb{V}^\Gamma \defines& \\
(\forall n, \sub_1, \sub_2: (&\sub_1, \sub_2) \in \mathbb{G}^\Gamma_n \Rightarrow (v_1[\sub_1], v_2[\sub_2]) \in \mathbb{V}_n)\\[5pt]
(e_1, e_2) \in \mathbb{E}^\Gamma \defines& \\
(\forall n, \sub_1, \sub_2: (&\sub_1, \sub_2) \in \mathbb{G}^\Gamma_n \implies (e_1[\sub_1], e_2[\sub_2]) \in \mathbb{E}_n)\\[5pt]
(\textit{exc}_1, \textit{exc}_2) \in \mathbb{X}^\Gamma \defines& \\
(\forall n, \sub_1, \sub_2: (&\sub_1, \sub_2) \in \mathbb{G}^\Gamma_n \implies (\textit{exc}_1[\sub_1], \textit{exc}_2[\sub_2]) \in \mathbb{X}_n)\\[5pt]
(r_1, r_2) \in \mathbb{R}^\Gamma \defines& \\
(\forall n, \sub_1, \sub_2: (&\sub_1, \sub_2) \in \mathbb{G}^\Gamma_n \implies (r_1[\sub_1], r_2[\sub_2]) \in \mathbb{R}_n)
\end{align*}
\end{definition}

\begin{figure*}

    \begin{prooftree}
    \hypo{(e_1, e_2) \in \mathbb{E}^{\Gamma \cup \{x_1, \dots, x_n\}}}
    \infer1{(\fun{x_1, \dots, x_n}{e}, \fun{x_1, \dots, x_n}{e_2}) \in \mathbb{E}^\Gamma}
    \end{prooftree}
    %
    \hfill
    %
    \begin{prooftree}
    \hypo{(e_1, e_1') \in \mathbb{E}^\Gamma}
    \hypo{(e_2, e_2') \in \mathbb{E}^\Gamma}
    \infer2{(\econs{e_1}{e_2}, \econs{e_1'}{e_2'}) \in \mathbb{E}^\Gamma}
    \end{prooftree}
    %
    \hfill
    %
    \begin{prooftree}
    \hypo{(e, e'), (e_1, e_1'), \dots, (e_n, e_n') \in \mathbb{E}^\Gamma}
    \infer1{(\apply{e}{e_1, \dots, e_n}, \apply{e'}{e_1', \dots, e_n'}) \in \mathbb{E}^\Gamma}
    \end{prooftree}

    \vspace{0.3cm}
    %
    \begin{prooftree}
    \hypo{(e_m, e_m'), (e_f, e_f'), (e_1, e_1'), \dots, (e_n, e_n') \in \mathbb{E}^\Gamma}
    \infer1{(\call{e_m}{e_f}{e_1, \dots, e_n}, \call{e_m'}{e_f'}{e_1', \dots, e_n'}) \in \mathbb{E}^\Gamma}
    \end{prooftree}
    %
    \hfill
    %
    \begin{prooftree}
    \hypo{(e_1^k, e_1^{k'}), (e_1^v, e_1^{v'}), \dots, (e_n^k, e_n^{k'}), (e_n^v, e_n^{v'}) \in \mathbb{E}^\Gamma}
    \infer1{(\emap{\mapitem{e_1^k}{e_1^v}, \dots, \mapitem{e_n^k}{e_n^v}}, \emap{\mapitem{e_1^{k'}}{e_1^{v'}}, \dots, \mapitem{e_n^{k'}}{e_n^{v'}}}) \in \mathbb{E}^\Gamma}
    \end{prooftree}

    \vspace{0.3cm}
    %
    \begin{prooftree}
    \hypo{(e_1, e_1'), \dots, (e_n, e_n') \in \mathbb{E}^\Gamma}
    \infer1{(\etuple{e_1, \dots, e_n}, \etuple{e_1', \dots, e_n'}) \in \mathbb{E}^\Gamma}
    \end{prooftree}
    \hfill
    %
    \begin{prooftree}
    \hypo{(e_1, e_1'), \dots, (e_n, e_n') \in \mathbb{E}^\Gamma}
    \hypo{a = a'}
    \infer2{(\primop{a}{e_1, \dots, e_n}, \primop{a'}{e_1', \dots, e_n'}) \in \mathbb{E}^\Gamma}
    \end{prooftree}
    %
    %
    \hfill
    %
    \begin{prooftree}
    \hypo{(e, e') \in \mathbb{E}^{\Gamma \cup \funsof{ext}}}
    \hypo{\equivext{\Gamma}{\textit{ext}}{\textit{ext}'}}
    \infer2{(\letrec{\textit{ext}}{e},\letrec{\textit{ext}'}{e'}) \in \mathbb{E}^\Gamma}
    \end{prooftree}
    
    \vspace{0.3cm}
    %
    \begin{prooftree}
    \hypo{(e_1, e_1'), \dots, (e_n, e_n') \in \mathbb{E}^\Gamma}
    \infer1{(\evalues{e_1, \dots, e_n}, \evalues{e_1', \dots, e_n'}) \in \mathbb{E}^\Gamma}
    \end{prooftree}
    %
    \hfill
    %
    \begin{prooftree}
    \hypo{(e_1, e_1') \in \mathbb{E}^\Gamma}
    \hypo{(e_2, e_2') \in \mathbb{E}^{\Gamma \cup \{x_1,\dots,x_n\}}}
    \infer2{(\elet{\varlist{x_1,\dots,x_n}}{e_1}{e_2}, \elet{\varlist{x_1,\dots,x_n}}{e_1'}{e_2'}) \in \mathbb{E}^\Gamma}
    \end{prooftree}
    \hfill
    %
    \begin{prooftree}
    \hypo{(e_1, e_1') \in \mathbb{E}^\Gamma}
    \hypo{(e_2, e_2') \in \mathbb{E}^\Gamma}
    \infer2{(\seq{e_1}{e_2}, \seq{e_1'}{e_2'}) \in \mathbb{E}^\Gamma}
    \end{prooftree}
    
    \vspace{0.3cm}
    %
    \begin{prooftree}
    \hypo{(e, e') \in \mathbb{E}^\Gamma}
    \hypo{\forall i \le n: (e^g_i, e^{g'}_i), (e^b_i, e^{b'}_i) \in \mathbb{E}^{\Gamma \cup \pvars{\textit{ps}_i}}}
    \infer2{(\case{e}{\clause{\textit{ps}_1}{e^g_1}{e^b_1}; \dots; \clause{\textit{ps}_n}{e^g_n}{e^b_n}}, \case{e'}{\clause{\textit{ps}_1}{e^{g'}_1}{e^{b'}_1}; \dots; \clause{\textit{ps}_n}{e^{g'}_n}{e^{b'}_n}}) \in \mathbb{E}^\Gamma}
    \end{prooftree}

    \vspace{0.3cm}
    \begin{prooftree}
    \hypo{(e_1, e_1') \in \mathbb{E}^\Gamma}
    \hypo{(e_2, e_2') \in \mathbb{E}^{\Gamma \cup \{x_1, \dots, x_k\}}}
    \hypo{(e_3, e_3') \in \mathbb{E}^{\Gamma \cup \{x_{k+1}, \dots, x_{k+n}\}}}
    \infer3{(\try{e_1}{\varlist{x_1, \dots, x_k}}{e_2}{\varlist{x_{k+1}, \dots, x_{k+n}}}{e_3}, \try{e_1'}{\varlist{x_1, \dots, x_k}}{e_2'}{\varlist{x_{k+1}, \dots, x_{k+n}}}{e_3'}) \in \mathbb{E}^\Gamma}
    \end{prooftree}
    
    \caption{Compatibility of expressions}
    \label{fig:expcompat}
\end{figure*}

\begin{figure*}
    %
    %
    \begin{prooftree}
    \hypo{x \in \Gamma}
    \infer1{(x, x) \in \mathbb{V}^\Gamma}
    \end{prooftree}
    \hfill
    \begin{prooftree}
    \hypo{f/k \in \Gamma}
    \infer1{(f/k, f/k) \in \mathbb{V}^\Gamma}
    \end{prooftree}
    \hfill
    \begin{prooftree}
    \infer0{(a, a) \in \mathbb{V}^\Gamma}
    \end{prooftree}
    \hfill
    \begin{prooftree}
    \infer0{(i, i) \in \mathbb{V}^\Gamma}
    \end{prooftree}
    \hfill
    %
    \begin{prooftree}
        \infer0{(\nil, \nil) \in \mathbb{V}^\Gamma}
    \end{prooftree}
    \hfill
    \begin{prooftree}
    \hypo{(v_1, v_1') \in \mathbb{V}^\Gamma}
    \hypo{(v_2, v_2') \in \mathbb{V}^\Gamma}
    \infer2{(\vcons{v_1}{v_2}, \vcons{v_1'}{v_2'}) \in \mathbb{V}^\Gamma}
    \end{prooftree}
    
    \vspace{0.3cm}
    
    \begin{prooftree}
    \hypo{(v_1, v_1'), \dots, (v_n, v_n') \in \mathbb{V}^\Gamma}
    \infer1{(\vtuple{v_1, \dots, v_n}, \etuple{v_1', \dots, v_n'}) \in \mathbb{V}^\Gamma}
    \end{prooftree}
    %
    \hfill
    %
    \begin{prooftree}
    \hypo{(v_1^k, v_1^{k'}), (v_1^v, v_1^{v'}), \dots, (v_n^k, v_n^{k'}), (v_n^v, v_n^{v'}) \in \mathbb{V}^\Gamma}
    \infer1{(\vmap{\mapitem{v_1^k}{v_1^v}, \dots, \mapitem{v_n^k}{v_n^v}}, \emap{\mapitem{v_1^{k'}}{v_1^{v'}}, \dots, \mapitem{v_n^{k'}}{v_n^{v'}}}) \in \mathbb{V}^\Gamma}
    \end{prooftree}
    \hfill
    \begin{prooftree}
    \hypo{(v_1, v_1'), \dots, (v_n, v_n') \in \mathbb{V}^\Gamma}
    \infer1{(\vvalues{v_1, \dots, v_n}, \vvalues{v_1', \dots, v_n'}) \in \mathbb{R}^\Gamma}
    \end{prooftree}

    \vspace{0.3cm}
    \begin{prooftree}
    \hypo{(e, e') \in \mathbb{E}^{\Gamma \cup \funsof{ext} \cup \{x_1, \dots, x_n\}}}
    \hypo{\equivext{\Gamma}{\textit{ext}}{\textit{ext}'}}
    \infer2{(\clos{\textit{ext}}{x_1, \dots, x_n}{e}, \clos{\textit{ext}'}{x_1, \dots, x_n}{e'}) \in \mathbb{V}^\Gamma}
    \end{prooftree}
    \hfill
    \begin{prooftree}
    \hypo{(v, v') \in \mathbb{V}^\Gamma}
    \infer1{(v, v') \in \mathbb{E}^\Gamma}
    \end{prooftree}
    \hfill
    \begin{prooftree}
    \hypo{(v_r, v_r') \in \mathbb{V}^\Gamma}
    \hypo{(v_d, v_d') \in \mathbb{V}^\Gamma}
    \infer2{(\exc{c}{v_r}{v_d}, \exc{c}{v_r}{v_d}) \in \mathbb{R}^\Gamma}
    \end{prooftree}
    \hfill
    \begin{prooftree}
    \hypo{(e, e') \in \mathbb{R}^\Gamma}
    \infer1{(e, e') \in \mathbb{E}^\Gamma}
    \end{prooftree}
    \hfill
    \begin{prooftree}
    \hypo{(e, e') \in \mathbb{E}^\Gamma}
    \infer1{(e, e') \in \mathbb{R}^\Gamma}
    \end{prooftree}
\caption{Other compatibility properties}
\label{fig:compatextra}
\end{figure*}

Next, we show the most important properties of the logical relations~\cite{wand2018contextual,pitts2010step,culpepper2017contextual}. The 
first one is the compatibility property, which is a form of congruence. For readability, we introduce the following definition.

\begin{definition}[Equivalence of function collections]
Two function collections ($\textit{ext}$ and $\textit{ext}'$) are related, if they bind the same names, and they are related 
element-wise. Two function definitions (denoted by $f/k = \fun{x_1, \dots, x_n}{e}$ and $f/k = \fun{x_1, \dots, x_n}{e'}$) of the 
collections $\textit{ext}$ and $\textit{ext}'$ are related, 
if they satisfy $(e, e') \in \mathbb{E}^{\Gamma \cup \funsof{\textit{ext}} \cup \{x_1, \dots, x_n\}}$. We use $\equivext{\Gamma}{\textit{ext}}{\textit{ext}'}$ to denote this property.
\end{definition}

\begin{theorem}[Expression Compatibility]\label{thm:compat}
  The logical relations satisfy the syntactical compatibility properties listed in \Cref{fig:expcompat}. Moreover, they also satisfy 
  extra compatibility properties listed in \Cref{fig:compatextra}.
\end{theorem}

Based on the previous theorem, we can prove that the logical relation for redexes and expressions coincide for related expressions.

\begin{corollary}[Equivalence of logical relations]
For all expressions $e_1, e_2$ and scopes $\Gamma$, $(e_1, e_2) \in \mathbb{E}^\Gamma \iff (e_1, e_2) \in \mathbb{R}^\Gamma$.
\end{corollary}

Another consequence of the compatibility theorem is the fundamental property of the relations, a form of reflexivity expressing that any expression (and similarly, any value, exception, redex or closing substitution) is indistinguishable from itself.

\begin{theorem}[Fundamental property]\label{thm:fundamental}
    For all scopes $\Gamma$ the following properties hold:
    \begin{itemize}
    \item For all expressions $e$, if $\expscoped{\Gamma}{e}$ then $(e, e) \in \mathbb{E}^\Gamma$;
    \item For all values $v$, if $\valscoped{\Gamma}{v}$ then $(v, v) \in \mathbb{V}^\Gamma$;
    \item For all exceptions $\textit{exc}$, if $\valscoped{\Gamma}{\textit{exc}}$ then $(\textit{exc}, \textit{exc}) \in \mathbb{X}^\Gamma$;
    \item For all redexes $r$, if $\valscoped{\Gamma}{r}$ then $(r, r) \in \mathbb{R}^\Gamma$;
    \item For all closing substitutions $\sub$, if $\subscoped{\Gamma}{\sub}{\emptyset}$ then for all step counters $n$, $(\sub, \sub) \in \mathbb{G}^\Gamma_n$ holds.
    \end{itemize}
\end{theorem}

Last but not least, another important property of the logical relations for values is that all related values should be equal by the built-in equality of 
(Core) Erlang (simulated by the auxiliary function $\resultof{\textit{call}(\texttt{'erlang'}, \texttt{'=='})}{v_1, v_2}$).

\begin{theorem}[Equivalent values are equal]
  For all values $v_1, v_2$, step counters $m$,
  \[
  (v_1, v_2) \in \mathbb{V}_m \implies \resultof{\textit{call}(\texttt{'erlang'}, \texttt{'=='})}{v_1, v_2} = \texttt{'true'}.
  \]
\end{theorem}

\subsection{CIU Equivalence}\label{sec:ciu}

Next, we introduce CIU (``closed instances of use'') preorder and equivalence.

\begin{definition}[CIU preorder] Two redexes are CIU equivalent if they both terminate or diverge when placed in arbitrary frame stacks.
\begin{align*}
  r_1 \ciupre r_2 \defines&\ \expscoped{\emptyset}{r_1}  \land \expscoped{\emptyset}{r_2}\  \land \\
  &(\forall K: \frameclosed{K} \land \term{K}{r_1} \implies \term{K}{r_2})\\
  r_1 \ciuequiv r_2 \defines&\ r_1 \ciupre r_2 \land r_2 \ciupre r_1
\end{align*}
We extend these concepts to open redexes with closing substitutions.
\begin{align*}
  r_1 \ciupre^\Gamma r_2 \defines&\ \forall \sub: \subscoped{\Gamma}{\sub}{\emptyset} \implies \subst{r_1}{\sub} \ciupre \subst{r_2}{\sub} \\
  r_1 \ciuequiv^\Gamma r_2 \defines&\ r_1 \ciupre^\Gamma r_2 \land r_2 \ciupre^\Gamma r_1
\end{align*}
\end{definition}

In most cases, it is simpler to prove redexes CIU equivalent, than using logical relations or contextual equivalence, because CIU 
equivalence involves reasoning with respect to a single frame stack instead of two related ones, or one syntactical context. One of 
the most important properties of CIU equivalence is that it coincides with logical relations on redexes.

\begin{theorem}[CIU coincides with the logical relations]\label{thm:ciulogrel}
For all redexes $r_1, r_2$, and scopes $\Gamma$, $r_1 \ciupre^\Gamma r_2$ if and only if $(r_1, r_2) \in \mathbb{R}^\Gamma$.
\end{theorem}


Another major property of CIU equivalence is that evaluating a redex results in an equivalent value sequence or exception.

\begin{corollary}[Redexes are equivalent to their results]\label{thm:ciueval}
For all closed redexes $r$, and results $\textit{res}$, if $\rewritesstarint{\idfs}{r}{\idfs}{\textit{res}}$, then $r \ciuequiv \textit{res}$.
\end{corollary}

Finally, we highlight one last property which expresses the fact that reasoning about termination of programs is sufficient for 
the final results to be equivalent.

\begin{theorem}[Termination is sufficient]\label{thm:ciuterm}
For all closed values $v_1, v_2$, if $v_1 \ciupre v_2$ then for all step indices $n$, $(v_1, v_2) \in \mathbb{V}_n$.
\end{theorem}

This theorem together with \Cref{thm:ciueval} and the transitivity of the equivalence relations (\Cref{sec:ctx}) means that
whenever two expressions are CIU equivalent, their values will be related by the logical relation for values, which expresses 
exactly what we expect from the behaviour of equivalent values.

\subsection{Contextual Equivalence}\label{sec:ctx}

Finally, we define contextual preorder and equivalence following the techniques of Wand et al.~\cite{wand2018contextual}.

\begin{definition}[Contextual preorder]\label{def:ctxpre}
We define the contextual preorder to be the largest family of relations $R^\Gamma$ that satisfy the following properties:
\begin{itemize}
	\item Adequacy: $(e_1, e_2) \in R^{\emptyset} \implies \term{\idfs}{e_1} \implies \term{\idfs}{e_2}$.
	\item Reflexivity: $(e, e) \in R^\Gamma$.
	\item Transitivity: $(e_1, e_2) \in R^\Gamma \land (e_2, e_3) \in R^\Gamma \implies (e_1, e_3) \in R^\Gamma$.
	\item Compatibility: $R^\Gamma$ satisfies the compatibility rules for every expression from~\Cref{fig:expcompat}.
\end{itemize}
\end{definition}

This definition is equivalent to the usual, syntax-based definition of contextual equivalence. We denote syntactical expression contexts 
with $C$ (where one of the subexpressions are replaced by a unique variable $\Box$), and use $\csubst{C}{e}$ to denote the substitution of 
$\Box$ with expression $e$ in context $C$.

\begin{definition}[Syntax-based contextual preorder and equivalence]\label{def:ctx}
\begin{align*}
e_1 \ctxpre^\Gamma e_2 \defines\ & \expscoped{\Gamma}{e_1} \land \expscoped{\Gamma}{e_2} \land (\forall (C : \textit{Context}):\\ \expscoped{\emptyset}{\csubst{C}{e_1}} &\land \expscoped{\emptyset}{\csubst{C}{e_2}} \implies \term{\idfs}{\csubst{C}{e_1}} \implies \term{\idfs}{\csubst{C}{e_2}}) \\
e_1 \ctxequiv^\Gamma e_2 \defines\ & e_1 \ctxpre^\Gamma e_2 \land e_2 \ctxpre^\Gamma e_1
\end{align*}
\end{definition}

The concepts above are all defined only for expressions, and not redexes. The reason for this is only expressions are syntactically 
valid Core Erlang expressions. There is no way to include an exception as a syntactical subexpression, because it is a semantical 
concept.

On the one hand, the previous definition of contextual equivalence expresses the correctness property of refactorings, that is 
replacing two equivalent expressions in any syntactical context preserves the behaviour, 
On the other hand, reasoning about contextual equivalence naively would require induction on the structure of the context. 
To tackle this issue, we proved that contextual equivalence coincides with CIU equivalence for expressions.

\begin{theorem}[CIU theorem]\label{thm:ctxciu}
For all expressions $e_1, e_2$, and scopes $\Gamma$, if and only if $e_1 \ctxpre^\Gamma e_2$, then $e_1 \ciupre^\Gamma e_2$.
\end{theorem}

As a consequence, we can state the following corollary on the connections between the equivalence concepts.

\begin{corollary}[Coincidence of equivalences]
For all expressions $e_1, e_2$, and scopes $\Gamma$, the following equivalences hold:
\[
  (e_1, e_2) \in \mathbb{E}^\Gamma \iff (e_1, e_2) \in \mathbb{R}^\Gamma \iff e_1 \ciupre^\Gamma e_2 \iff e_1 \ctxpre^\Gamma e_2
\]
\end{corollary}

\subsection{Refactoring Correctness}\label{sec:examples}

With contextual equivalence, we can express the correctness property of refactorings.
A local refactoring which replaces a subexpression $e$ with $e'$ is correct, if $e \ctxequiv^\Gamma e'$ (supposing that $e$ and $e'$ 
contain the free variables in $\Gamma$). Proving the contextual equivalence implies that the behaviour of the entire context (i.e., a whole program) does not 
change when the two equivalent expressions are replaced.

As mentioned previously, reasoning about contextual equivalence is not simple in most cases, thus we prove CIU equivalence of expressions instead, and 
use \Cref{thm:ctxciu} to establish the contextual equivalence. We proved the correctness of simple Erlang refactorings (two examples are 
\Cref{fig:refactor1} and \ref{fig:refactor2}).

\begin{figure}[ht]
\centering
\begin{subfigure}{0.24\textwidth}
\begin{lstlisting}
  case $e_1$ of true -> $e_2$;
             _ -> $e_3$
  end
\end{lstlisting}
\end{subfigure}
{\LARGE$\rightarrow$}
\begin{subfigure}{0.18\textwidth}
\begin{lstlisting}
  if $e_1$ -> $e_2$;
     true -> $e_3$
  end
\end{lstlisting}
\end{subfigure}
\caption{Expression refactoring example}
\label{fig:refactor2}
\end{figure}

To argue about these transformations, first we translated these programs to Core Erlang with the standard Erlang/OTP compiler 
(which we handled as trusted component for the proving process). Next, we encoded the Core Erlang programs in the Coq formalisation 
and proved their equivalence. We based the equivalence proofs on the termination of the expressions, that is we inspected all possible 
termination paths for one and proved the termination of the other expression, based on the properties we obtained from the first 
evaluation. For this step, the inductive definition of the frame stack termination proved to be extremely useful as its rules are (mostly) syntax-driven and most of them are free of side conditions.


\section{Discussion}\label{sec:disc}

All results presented here are formalised in the Coq proof management system~\cite{coreerlang}. 
In this section, we highlight a number of challenges we faced during the implementation.

\paragraph{Syntax} Initially, we considered two other approaches to formalise the syntax: (1) values are completely separated from 
expressions and (2) there are only expressions and a judgement which determines whether an expression is a value (i.e., is in normal form).
The issue with 
(1) is applying substitutions requiring the substituted values to be transformed to expressions, leading to loss of information 
in case of closures (the list of recursive definitions would simply be lost).
The disadvantage of (2) is that the value judgement relation needs to be 
used in most of the rules of the semantics to ensure determinism, which on the other hand leads to more proof steps about the evaluation.
However, the 
approach presented in \Cref{sec:syntax} does not come without drawbacks either: using mutually inductive types in Coq leads to more complicated induction principles 
(which had to be defined manually) and theorem statements about the syntax.

\paragraph{Semantics} The main advantage of the frame stack semantics is that most of the rules can be applied in a syntax-directed way, 
which significantly simplifies proving evaluation. The notion of parameter list frames was motivated by the implementation to avoid the 
duplication of frames, reduction rules, theorems for similar language elements: tuples, maps, built-in function calls, primitive operations, 
and function applications. With parameter list frames, these features can be handled in a unified way. This notion was also used 
for reduction contexts by Fredlund~\cite{fredlund2001framework}.


\paragraph{Logical relations} We formalised logical relations with definitions that are parametrised by the step-indexed value relation ($\mathbb{V}_n$) instead of mutually inductive types, following the footsteps of Wand et al.~\cite{wand2018contextual}. This way, we also 
avoided the strict positivity checks of Coq for inductive types.

\paragraph{Induction principles} Further interesting points in the formalisation are induction principles. We highlight the induction principle 
for the logical relation on values ($\mathbb{V}_n$). While using induction on the logical relation, only relevant cases that contain related values 
need to be proved, and the rest will be filtered out (e.g., we do not need to derive a contradiction from premises such as $(i, \nil) \in \mathbb{V}_n$).

\section{Related Work}\label{sec:work}

Our previous work and the result here on Core Erlang is based on the language specification~\cite{carlsson2000core} and related research. 
The most influential ones are reversible semantics for Erlang~\cite{lanese2018theory,lanese2018cauder,nishida2016reversible}, a 
framework for reasoning about Erlang~\cite{fredlund2001framework}, symbolic execution~\cite{vidal2014towards}, and abstraction 
and model checking~\cite{neuhausser2007abstraction}.

In related work, CIU equivalence~\cite{ahmed2006stepindexed,birkedal2013stepindexed,craig2018triangulating,culpepper2017contextual,wand2018contextual,gordon1999compilation,mason1991equivalence} and logical relations (either type-indexed~\cite{sumii2005logical,pitts2000operational} or step-(and type-) indexed~\cite{ahmed2006stepindexed,wand2018contextual,pitts2010step,culpepper2017contextual}) were successfully 
applied for a wide variety of languages 
(e.g., different variants of lambda calculi, imperative languages). Most of the related works---that define CIU equivalence---use a 
continuation-style semantics, similarly to our case where the frame stack can be seen as the continuation.
The novelty of our work lies with the choice of the language, the extent of the language elements formalised, and the machine-checked 
implementation.

In the related literature, there are other options to formalise program equivalence. The most simple notion is behavioural 
equivalence~\cite{pierce2010software} which is based on syntactical equality of the evaluation results. Another approach is 
using bisimulations~\cite{simpson2019behavioural,pitts1997operationally,abramsky1990thelazy,lanese2019playing} which are relations 
between programs preserved by the reduction steps.

\section{Conclusion and Future Work}\label{sec:conclusion}

In this paper, we defined a formal syntax and a frame stack semantics for sequential Core Erlang. Thereafter, we presented a number of properties 
of this semantics, and defined three expression equivalence concepts (based on logical relations, CIU equivalence, and contextual equivalence). 
We showed that these termination-based equivalences are sufficient to ensure the final results of equivalent programs to be behaviourally 
indistinguishable. Moreover, we also showed that these three equivalence concepts coincide for (Core Erlang) expressions.

In the short term future, we are going to extensively validate the frame stack semantics presented here by showing its equivalence with
the validated big-step semantics in our previous work~\cite{bereczky2021validation}. 
In the longer term, we are going to combine this work with related research on the concurrent subset of 
Core Erlang~\cite{lanese2019playing,bereczky2022formalisation}. Moreover, we 
also plan to investigate more complex (non-local) refactorings (both for the sequential and concurrent sublanguage) based on the semantics 
and equivalence concepts defined here.

\begin{acks}
Supported by the ÚNKP-22-3 New National Excellence Program of the Ministry for Culture and
Innovation from the source of the National Research, Development and Innovation Fund.
\end{acks}

\bibliographystyle{ACM-Reference-Format}
\bibliography{bibfile}


\end{document}
\endinput